# Sunward Streaming $^3$He-rich SEP Events Observed by Solar Orbiter and Parker Solar Probe during Perihelion Passage

Samuel T. Hart[1], George C. Ho[1,5], Michael Terres[2], Gabriel C. Muro[3], Robert C. Allen[1], Maher A. Dayeh[1,5], Radoslav Bučík[1], Glenn M. Mason[4], Athanasios Kouloumvakos[4], Abdullah A. Shmies[5,1]

[1]Southwest Research Institute, San Antonio, TX 78238, USA
[2] Center for Astrophysics—Harvard & Smithsonian, Cambridge, MA 02138, USA
[3] Space Radiation Laboratory, California Institute of Technology, Pasadena, CA 91125, USA
[4]Applied Physics Laboratory, Johns Hopkins University, Laurel, MD 20723, USA
[5]The University of Texas at San Antonio, San Antonio, TX 78249, USA

## Abstract

We report on two $^3$He-rich solar energetic particle (SEP) events observed by Solar Orbiter (SO) and Parker Solar Probe (PSP) during the April 1 – 4, 2024 conjunction when both spacecraft are within 0.3 AU near their respective perihelion passage. The two $^3$He-rich SEP events, originating from active region (AR) 13615, exhibit two key anomalies: (1) sunward streaming SEPs and (2) SEP travel path lengths exceeding 2 – 8 times the nominal Parker spiral expectations. Remote observations suggest these SEPs have been redirected by a slow coronal mass ejection (CME) that also originated from AR 13615 two days prior on March 30, 2024 at 21:04 UT. Using the near-Sun CME speed and width measurements, we estimate its size and location at the onset time of the first $^3$He-rich SEP event. Based on our estimates, SEPs propagating around the ICME front have travelled between 0.76 – 0.95 AU when they arrive at SO, increasing to 0.94 – 1.1 AU at PSP, consistent with our observed SEP arrival times. These findings constitute the first multi-spacecraft observation of sunward streaming $^3$He-rich SEPs. We discuss the implications of this phenomenon on $^3$He-rich seed material and the rare widespread $^3$He-rich SEP events (see §5.4).

# 1. Introduction

Solar energetic particle (SEP) events are an excellent tool to understand particle acceleration and transport processes in solar and stellar atmospheres. Among them, $^3$He-rich SEP events are notable for their extreme enrichments in $^3$He relative to $^4$He and moderate enrichments in Ne – Fe relative to O compared to the solar wind (Reames 1999, Mason 2007, Reames 2021, Kouloumvakos et al. 2025). Furthermore, $^3$He-rich SEPs are entirely accelerated in the lower solar corona and directly released into interplanetary space, making them useful probes for particle acceleration within solar active regions. Though progress has been made characterizing these peculiar events, many questions remain regarding their origins, acceleration, transport, and their role in supplying a pre-accelerated "seed" population for large, gradual SEP events (see Bučík 2020, Ho et al. 2022, Laming & Kuroda 2023, Mason et al. 2023a, Fitzmaurice et al. 2024, Bučík et al. 2025a and references therein).

$^3$He-rich SEP events are closely linked with type III radio bursts (Reames & Stone 1986, Mason et al. 2023b) but are weakly correlated with large coronal mass ejections (CMEs) and type II radio bursts (Kahler et al. 1985). Recent solar source observations show $^3$He-rich SEP events are not associated with strong X-ray flares, but are instead associated with ascending extreme ultraviolet (EUV) jets (Wang et al. 2006, Nitta et al. 2015, Bučík 2020) suggesting these events are driven by interchange reconnection along open/closed magnetic field boundaries (Shimojo & Shibata 2000, Wang et al. 2006). Once in interplanetary space, $^3$He-rich SEPs exhibit limited spatial extent, spanning typically a few 10s of degrees in heliospheric longitude (Ho et al. 2024). However, on rare occasions, widespread distributions of $^3$He-rich SEPs over 120º in heliospheric longitude have been observed (Wiedenbeck et al. 2013, 2020).

Through solar cycle 25, Solar Orbiter (SO; Müller et al. 2020) observations have revealed new insights into the origins and properties of $^3$He-rich SEP events. Inner heliospheric observations show numerous events exhibit atypical compositions, including: (1) events with the largest recorded $^3$He/$^4$He relative abundance enhancements (Kouloumvakos et al. 2025), (2) events with atypical heavy ion abundances (Bučík et al. 2025b) that are not sorted by mass (see Mason et al. 2004), and (3) events with exceptionally large ultra-heavy ion enrichments (>80 amu; Mason et al. 2023a). One important revelation from SO and Parker Solar Probe (PSP; Fox et al. 2016) observations is some SEP events exhibit unusually long travel path lengths 2 – 4 times longer than the nominal Parker spiral magnetic field (Wimmer-Schweingruber et al. 2023, Xu et al. 2024). Long path lengths at 1 AU are often attributed to interplanetary magnetic field (IMF) random walk, or field line meandering (Laitinen & Dalla 2019, Chhiber et al. 2021). However, in the inner heliosphere, the path lengths are too long to be attributed solely to field line meandering. For example, Wimmer-Schweingruber et al. (2023) report on a $^3$He-rich SEP event measured *in situ* by SO during the passage of an interplanetary CME (ICME) flux rope, and the authors conclude that the SEPs traveled along the broad ICME flux rope, yielding the longer path length.

In this work, we investigate a pair of $^3$He-rich SEP events observed by both SO and PSP that exhibit anomalously long path lengths (over 2 times nominal Parker spiral) and a bulk SEP flow *toward* the Sun rather than away from it. Sunward streaming SEPs are commonly observed in shock-accelerated transient events (Starkey et al. 2024, Wei et al. 2024, Allen et al. 2025) because shocks continuously accelerate and release SEPs as they propagate through interplanetary space. However, $^3$He-rich SEP events are accelerated entirely in the solar corona, suggesting these $^3$He-rich SEPs have been ushered back toward their solar origin. We present ion anisotropies and energy spectra of this pair of $^3$He-rich events at SO and PSP and identify their solar source regions. We investigate the cause of sunward streaming SEPs and provide evidence indicating these $^3$He-rich SEPs were redirected by a slow-moving (sub-400 km/s) ICME, and these SEPs are observed only after they have fully wrapped around the ICME. Finally, we discuss this work's implications on our understanding of widespread ³He-rich SEP events as well as our understanding of remnant ³He-rich SEPs in the solar corona.

## 2. Data and Instrumentation

Our study primarily uses energetic particle data from the Solar Isotope Spectrometer (SIS) within the Energetic Particle Detector suite (Rodriguez-Pacheco et al. 2020, Wimmer-Schweingruber et al. 2021) on board SO. SIS consists of two time-of-flight – energy (TOF-E) telescopes to measure incident ion mass with high precision between 0.04 – 10 MeV/nuc. SIS has a mass resolution of $\sigma_m/m < 0.02$, enabling clear peak separation of ³He, ⁴He, and the C – Fe heavy ions. The look directions of these two telescopes are separated by 130º in the orbital plane. SIS-A, the sunward facing telescope, nominally points 30º from the Sun, while SIS-B typically points 160º from the Sun. We also use He intensity measurements from Energetic Particle Investigation's high energy *dE/dx – E* telescopes (EPI-Hi) within the IS☉IS suite (McComas et al. 2016) on board PSP. LET consists of three telescopes: LET-A and LET-B point in opposite directions with LET-A typically pointing 45º from the Sun in the ram direction, and LET-C generally points 135º from the Sun in the ram direction. As with SIS, all three LET telescopes have a central look direction along the spacecraft's orbital plane. However, these look directions are subject to change during spacecraft roll operations, and we have verified no such operations occur during this period.

For solar wind electron, ion, suprathermal electron Strahl measurements observed by SO, we use the Proton-Alpha System (PAS) and Electrostatic Analyzer System (EAS) within the Solar Wind Analyzer (SWA; Owen et al. 2020) suite. For PSP, we use the Solar Wind Electrons, Alphas, and Protons (SWEAP; Kasper et al. 2016) investigation. IMF measurements are obtained using the fluxgate magnetometer on board SO (Horbury et al. 2020) and search coil magnetometer on board PSP (FIELDS; Bale et al. 2016). For solar source images, we utilize the Extreme Ultraviolet Imager (EUI; Rochus et al. 2020) on board SO, and we use the Large-Angle Spectroscopic Coronagraph (LASCO; Brueckner et al. 1995) on board the Solar and Heliospheric Observatory spacecraft (SOHO; Domingo et al. 1995) for white-light coronagraph images. Finally, the ephemerides are obtained using the SPICE ancillary information system (Acton 1996). All *in situ* data products used in this work are publicly available on the Coordinate Data Analysis Website (CDAWeb), and solar images are available at the Solar Orbiter Archive (SOAR).

### 3. *In Situ* Observations

*3.1 Ephemeris Overview*

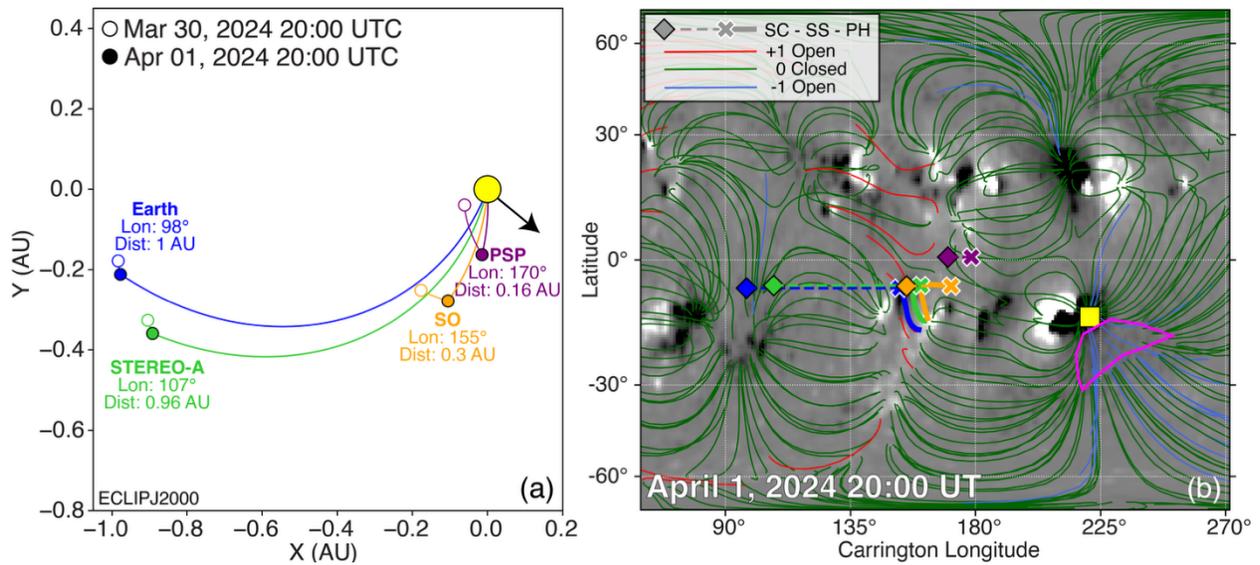

**Figure 1.** (a) Ephemerides of the relevant satellites on March 30, 2024 20:00 UT (open circles) through April 1, 2024 20:00 UT (filled circles). The source longitude of the AR producing the $^3$He-rich SEP events is shown as a black arrow. (b) PFSS solution at the time of the first $^3$He-rich SEP event, showing each spacecraft to source surface to photosphere connection (SC-SS-PH) and the polarity of the open and closed field lines. Spacecraft colors are consistent across the two panels. AR 13615 and the neighboring CH are shown in yellow and magenta.

Figure 1a shows the spacecraft positions projected on the ecliptic of J2000 (ECLIPJ2000) from March 30, 2024 20:00 UT to April 1, 2024 20:00 UT, the latter time coinciding with the initial release of $^3$He-rich SEPs from AR 13615. During this period, PSP is located at 0.07 – 0.16 AU on an outbound trajectory following its 19$^{th}$ perihelion, and SO is at ~0.3 AU just prior to its perihelion. STEREO-A and ACE are separated by only 9º in heliographic longitude. We show nominal Parker spiral field lines for each spacecraft using the instantaneous solar wind speed measured *in situ* on April 1, 2024 20:00 UT. Figure 1b shows the potential field source surface (PFSS; Wang & Sheeley 1992) solution to the 10$^{th}$ realization of the Air Force Data Assimilative Photospheric Flux Transport (ADAPT; Arge et al. 2010, Poirier et al. 2021) synoptic magnetogram provided every two hours. We trace each spacecraft's ideal Parker spiral to the source surface at 2.5 $R_S$, and the PFSS model solution connects each spacecraft's source surface connection to its photospheric footpoint. All four spacecraft have similar footpoint connectivity on and around April 1, 2024, but none of them are connected to AR 13615 separated by approximately 50º in heliospheric longitude from the spacecraft's footpoints. On the leading edge of AR 13615, there is a low-latitude coronal hole (see Appendix A), indicating the $^3$He-rich SEP events likely originate from the AR–CH boundary.

*3.2 Measurement Overview*

*3.2.1 Solar Orbiter*

In Figure 2, we present an overview of SO observations during the two $^3$He-rich SEP events. Figure 2a shows the radio observations from PSP/FIELDS. While we observe many type III radio bursts during this three-day stretch, the two type III radio bursts associated with our two $^3$He-rich SEP events occur at April 1, 2024 19:58 UT and April 3, 2024 02:05 UT, respectively, in

Sunward Streaming $^3$He-rich SEPs                                                                     6

conjunction with moderate enhancements in the >10 keV electron intensities (not shown). Figures 2b & 2c show the 15-minute 200 keV/nuc – 10 MeV/nuc $^3$He differential energy spectrograms measured by the sunward facing telescope (SIS-A) and the antisunward facing telescope (SIS-B), and Figure 2d & 2e show the same plots for the C – Fe heavy ions, which have been summed together for better statistics and multiplied by energy (MeV/nuc). After the type III radio bursts, we observe two distinct velocity dispersion profiles corresponding to the arrival of the two $^3$He-rich SEP events. The first $^3$He-rich SEP event is more intense, with observable enhancements in SEP fluxes up to ~10 MeV/nuc. The second event occurs ~30 hours later during the decay phase of the first event but is still clearly observed. Qualitatively, SEP fluxes just after event onsets are slightly higher in the antisunward-facing SIS-B compared with sunward-facing SIS-A, particularly above 1 MeV/nuc. We overplot the velocity dispersion profile for scatter-free SEPs travelling from the Sun to Solar Orbiter along the nominal Parker spiral (0.3 AU) as a dashed white curve and a much longer 1 AU path length as a white dot-dashed curve. The observed dispersion profiles are inconsistent with the nominal Parker spiral path, but rather both events align more closely with the 1 AU travel path length, over 3x greater than expected under nominal IMF conditions. Figures 2f – 2g show the solar wind conditions at SO during the same period. From the onset of the first SEP event, the solar wind speed remains relatively stable with no long-lasting velocity spikes or troughs through the majority of the second SEP event. The density similarly shows no drastic changes during this period. During the first SEP event, the IMF magnitude exhibits minimal change, and the radial component of the IMF, $B_R$, remains positive and is the dominant IMF component as expected at 0.3 AU. During the second event, the IMF magnitude increases slightly and is no longer consistently radial, but there are no obvious signs of large-scale structures. The pitch angle distributions of the 150 eV Strahl electrons are shown in Figure 2h, exhibiting unimodal distributions streaming away from the Sun. Our observations contrast those of Wimmer-Schweingruber et al. (2023), where the event is clearly observed during the passage of a well-defined flux rope.

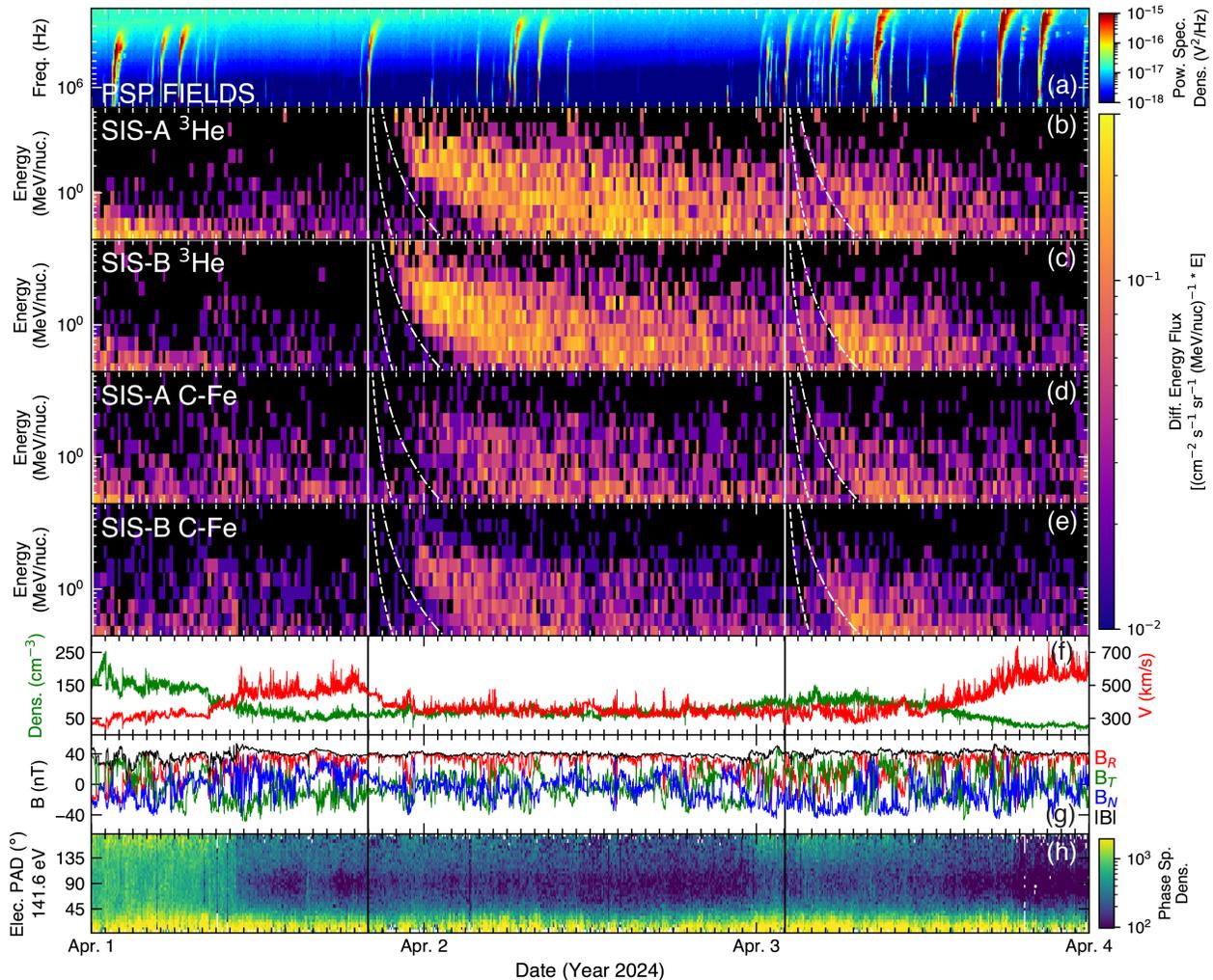

**Figure 2.** Overview of Solar Orbiter observations during the two sunward streaming $^3$He-rich SEP events. (a) PSP/FIELDS radio burst observations. (b) SIS-A $^3$He 0.3 – 8 MeV/nuc differential energy flux spectrogram. The type III radio burst associated with the two events are shown as vertical white lines, and the curved white lines show the velocity dispersion for particles travelling 0.3 AU (dashed) and 1 AU (dot-dashed). (c) Same as (b) but for SIS-B. (d – e) Same as (b – c) but for C – Fe heavy ions. (f) SWA/PAS solar wind density (green) and speed (red). (g) IMF magnitude and components in RTN coordinates. (h) SWA/EAS 142 eV electron strahl pitch angle distributions.

Sunward Streaming $^3$He-rich SEPs                                                                                          8

*3.2.2 Parker Solar Probe*

During this same period, PSP is on an outbound trajectory following its 19th perihelion. As illustrated in Figure 1, PSP and SO are nearly magnetically aligned, and we provide an overview of PSP/EPI-Hi observations in Figure 3. The 2D He (combined $^3$He and $^4$He) differential energy spectrograms for LET-A, B, and C are shown in Figures 3b – 3d. H and C – Fe enhancements are not observed by PSP during this period, H due to overwhelming background and C – Fe due to intensities remaining below the detection threshold. Additionally, neither of the two events are observed by PSP/EPI-Lo. Nevertheless, event 1 is clearly observed in all three LET telescopes. We observe no clear enhancement associated with event 2 despite close proximity to SO, and we attribute this discrepancy to event 2 being less intense above a few MeV/nuc (see Figure 3). The vertical white line shows the onset of the type III burst, and the dashed and dot-dashed white curves are the 0.16 AU and 1 AU velocity dispersion lines, respectively. Just as with SO, the dispersion profiles are consistent with an elongated SEP travel path length. Figure 3e and 3f show the solar wind density and IMF magnitude decreasing due to PSP's outbound trajectory, with the solar wind speed gradually increasing over the duration. Figure 3g shows the 134 eV Strahl electron pitch angle distribution. Overall, the solar wind properties at PSP are not indicative of a passing large-scale structure.

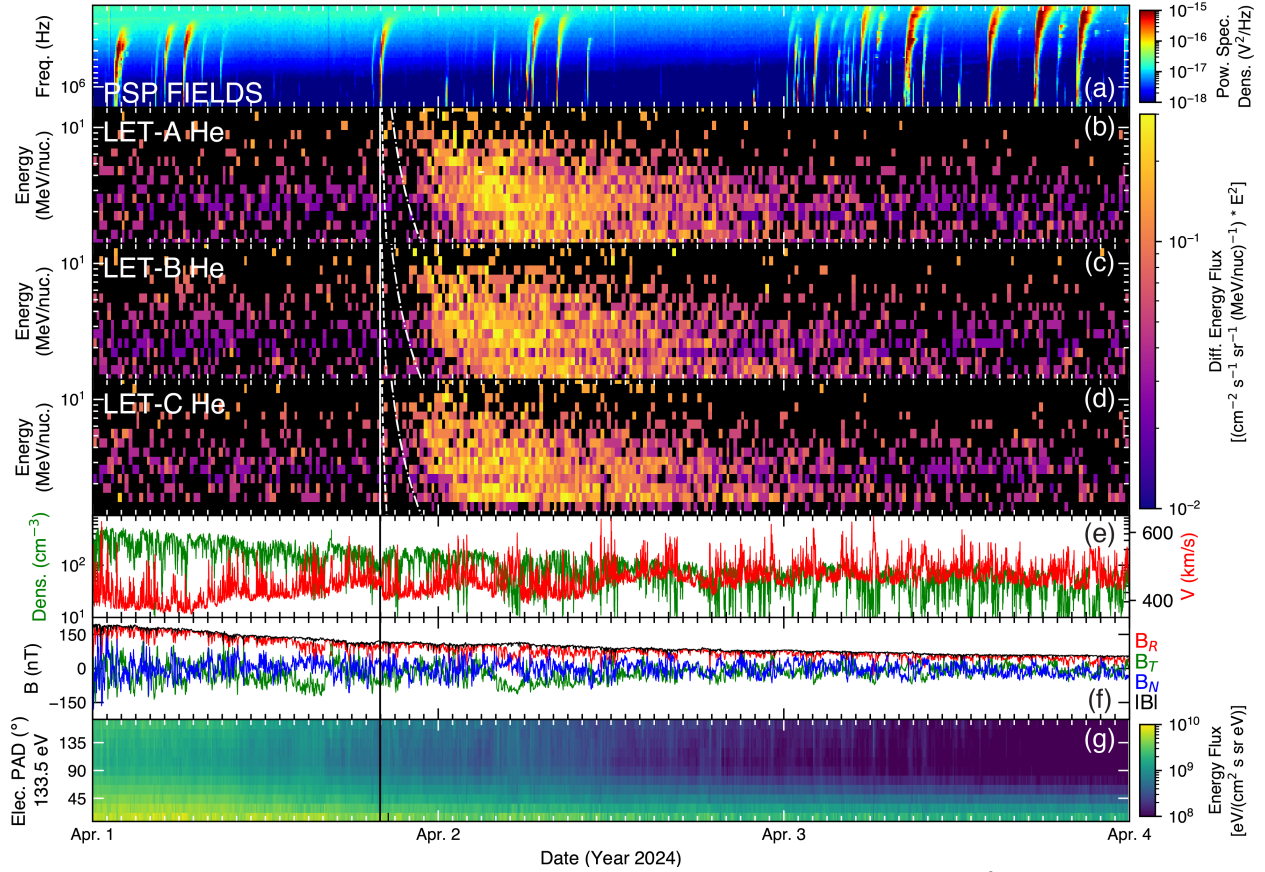

**Figure 3.** Overview of PSP observations during the two sunward streaming $^3$He-rich SEP events. (a) Same as Figure 2a. (b) LET-A He 1 – 15 MeV/nuc differential energy flux spectrogram. The vertical white line indicates type III burst onset time. The dashed and dot-dashed white curves are the velocity dispersion profiles for a 0.16 AU and 1 AU path length, respectively. (c – d) Same as (b), but for LET-B and LET-C. (e) PSP/SWEAP solar wind density (green) and speed (red). (f) IMF magnitude and components. (g) PSP/SWEAP 134 eV electron Strahl pitch angle distributions.

### 3.2.3 Measured SEP Path Lengths

To measure the SEP path lengths, we adopt a simple yet effective approach. We define the type III radio burst times as the solar particle release times, accounting for light travel time, and we calculate the path length based on the SEP arrival times in a single energy bin. While more rigorous velocity dispersion analysis methods exist (Kouloumvakos et al. 2015, Zhao et al. 2019, Palmroos et al. 2025, Shmies et al. 2026), our approach enables straightforward comparisons across two spatially separated spacecraft. For this analysis, we use 1.47 MeV/nuc for SO and 1.68 MeV/nuc for PSP because SO/SIS and PSP/LET energy bins overlap, and it provides sufficient statistics for easier identification of SEP onset times. Uncertainties in the path lengths are calculated based on the 15-minute cadence of the measurements. For event 1, we measure an SEP arrival time of 22:23 ± 00:07.5 UT for SO and 23:08 ± 00:07.5 UT for PSP corresponding to a travel time from the Sun of 145 minutes and 190 minutes, respectively. From there, we calculate a total path length of 0.98 ± 0.05 AU and 1.37 ± 0.05 AU. Given that SO and PSP are located at 0.3 AU and 0.16 AU, these paths are 3.3x and 8.6x greater than nominal Parker spiral values. The arrival time of SEPs at SO for event 2 is 04:08 ± 00:07 UT on April 3, 2024 corresponding to a travel time of 123 minutes and a path length of 0.83 ± 0.05, a 16% decrease from that of event 1, but still nearly three times nominal values.

*3.3 Energy Spectra & Anisotropy*

*3.3.1 Solar Orbiter*

$^3$He-rich SEP events are typically divided into two phases: the dispersive phase (or rise phase) and the decay phase. Typically, SEPs in the dispersive phase stream away from the Sun, producing strong anisotropies lasting several hours. Over time, SEPs experience are pitch-angle scattered toward an isotropic distribution (Mason et al. 2021). For SO, we define the dispersive phase of each event to be 8 hours after onset with an energy dependent onset time following a 1 AU travel path length. The decay phase for event 1 persists until the dispersive phase of event 2, and the decay phase of event 2 lasts 14 hours until intensities fall to background levels.

Because SEP pitch-angle scattering occurs in the solar wind frame, the isotropic distributions of the decay phase are relative to the solar wind frame. In the spacecraft frame, however, SEPs exhibit a slight anisotropy favoring downstream propagation (in the direction of solar wind flow) due to the Compton-Getting effect (Compton-Getting 1935, Ipavich 1974). For SO/SIS, isotropic distributions in the solar wind frame yield SIS-A/SIS-B flux ratios above unity, increasing at lower energies as the solar wind speed is a non-negligible fraction of the SEP speed ($v_{SW}/v_{SEP} \sim 10\%$ for 100 keV/nuc SEPs). We convert the SIS-A and SIS-B energy spectra from the spacecraft frame to the solar wind frame following Zirnstein et al. (2021) and Henderson et al.

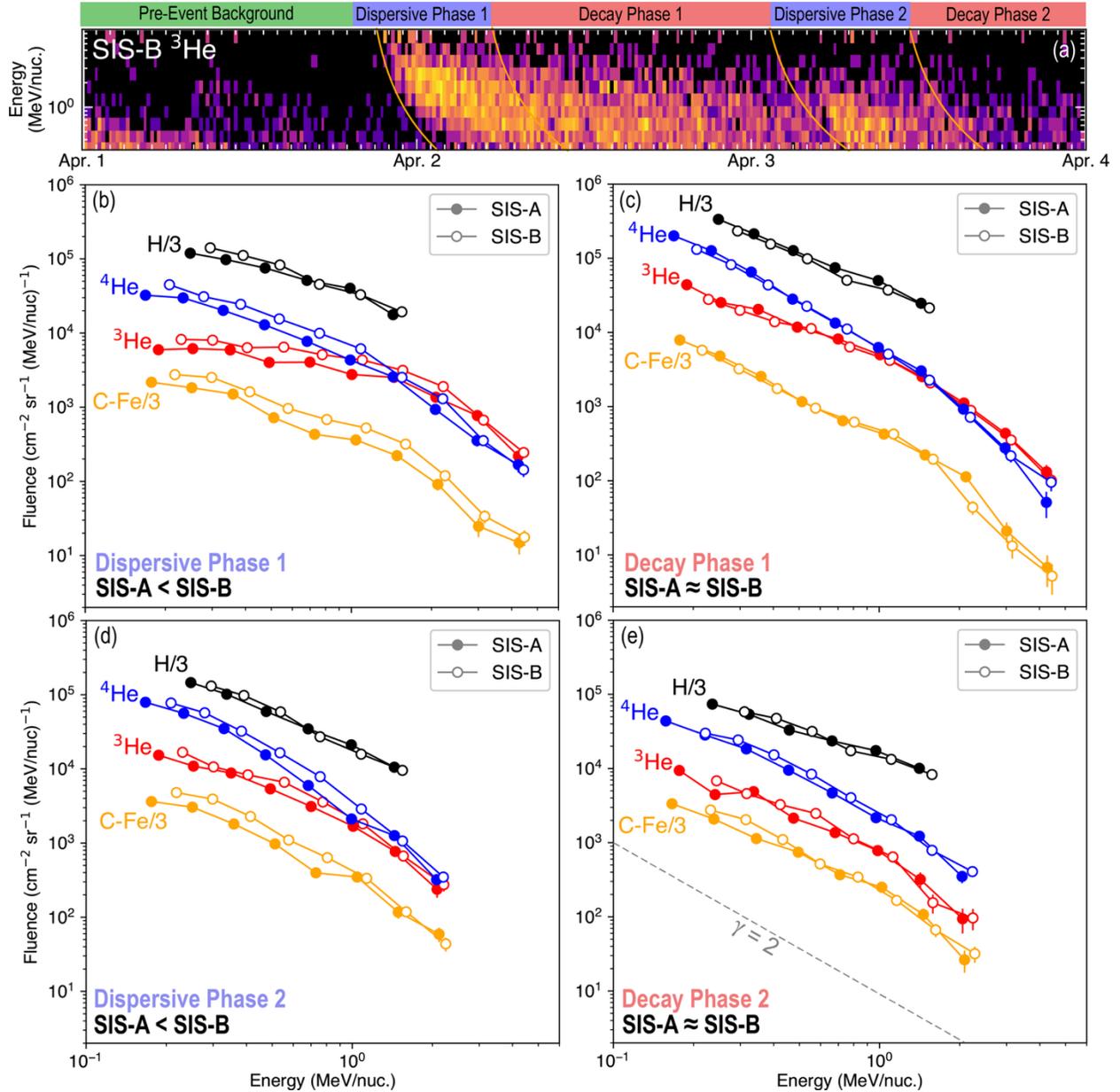

**Figure 4.** (a) Same as Figure 2c, but with each phase defined. Energy spectra of the dispersive phase (b, d) and decay phase (c, e) of the two $^3$He-rich SEP events observed by Solar Orbiter. In both dispersive phases, the SIS-B intensity (open circles) is greater than the SIS-A intensity (closed circles), indicating a bulk SEP flow toward the Sun at event onset. Both decay phases show isotropic distributions. H and C – Fe have been divided by a factor of 3 for clarity.

(2025), and the resulting Compton-Getting corrected energy spectra for each phase are shown in Figure 4. Figure 4a shows Figure 2c, but we have marked the different phases used to generate the energy spectra. Figure 4b shows the energy spectra of H, $^3$He, $^4$He, and C – Fe fluences during the dispersive phase of event 1. SIS-B (open circles) consistently measures larger fluences compared to SIS-A (filled circles) across the 0.1 – 5 MeV/nuc energy range, indicating net sunward flow of the $^3$He-rich SEPs. By contrast, during the decay phase of event 1 (Figure 4c), the SIS-A and SIS-B energy spectra nearly coincide at all energies consistent with an isotropic distribution. Event 2 is less intense, but exhibits the same sunward anisotropies during the dispersive phase (Figure 4d) and isotropic distribution during the decay phase (Figure 4e). Proton fluences above 2 MeV/nuc are excluded from our analysis due to exceptionally low detection efficiencies. Notably, the spectral shapes and relative abundances observed in these events are consistent with previously observed $^3$He-rich SEP events: (1) $^3$He exhibits harder low-energy spectra compared to $^4$He (Figure 4b, d; see Mason et al. 2002, Bučík et al. 2018, Hart et al. 2024), (2) nominal $^3$He/$^4$He and Fe/O abundance ratios across the energy range (Hart et al. 2022, Kouloumvakos et al. 2025), and (3) the power-law spectra with exponential rollover is commonly observed in $^3$He-rich SEP events (Hart et al. 2024, Mason et al. 2024). The only two obvious abnormalities present in the *in situ* observations are the exceptionally long path lengths and the bulk sunward streaming of the SEPs.

To further quantify the bulk sunward flow of these two $^3$He-rich SEP events, we calculate the first-order anisotropy as a function of time and energy. Anisotropy can be computed using a variety of methods. For measurements with high angular resolution, anisotropy can be calculated using fits to higher-order Legendre Polynomials (see Carcaboso et al. 2020 and references therein). When angular resolution is limited, as is the case for SO/SIS, anisotropy measurements are reduced to their first order (Dresing et al. 2014, Wei et al. 2024). We calculate the first-order anisotropy of each species by comparing the SEP fluxes between the two SIS telescopes at each energy bin using the following expression:

$$I_1 = \frac{J(A) - J(B)}{J(A) + J(B)} \quad (1)$$

where $I_1$ is the first-order anisotropy, and $J(A)$ and $J(B)$ are the CG-corrected differential fluxes measured by SIS-A and SIS-B at some energy bin, $E$. Positive $I_1$ values indicate bulk flow away from the Sun, negative $I_1$ values indicate bulk flow toward the Sun, and isotropic distributions yield $I_1 = 0$. As the Compton-Getting (CG)-correction adjusts the central energy of each energy bin to the solar wind frame ($E$), we logarithmically interpolate the CG-corrected fluxes ($J$) to the original energy bins ($E_O$) with the following expression:

$$J_O = J\left(\frac{E_O}{E}\right)^{-\gamma} \tag{2}$$

where $J_O$ is the interpolated flux, and $\gamma$ is the spectral index. The spectral index, $\gamma$, is derived by:

$$\gamma = -\frac{\ln(J_l/J_u)}{\ln(E_l/E_u)} \tag{3}$$

where $u$ and $l$ denote upper and lower flux and energy values (Dayeh et al. 2012). This interpolation ensures that the the CG-corrected fluxes are compared at equivalent energies.

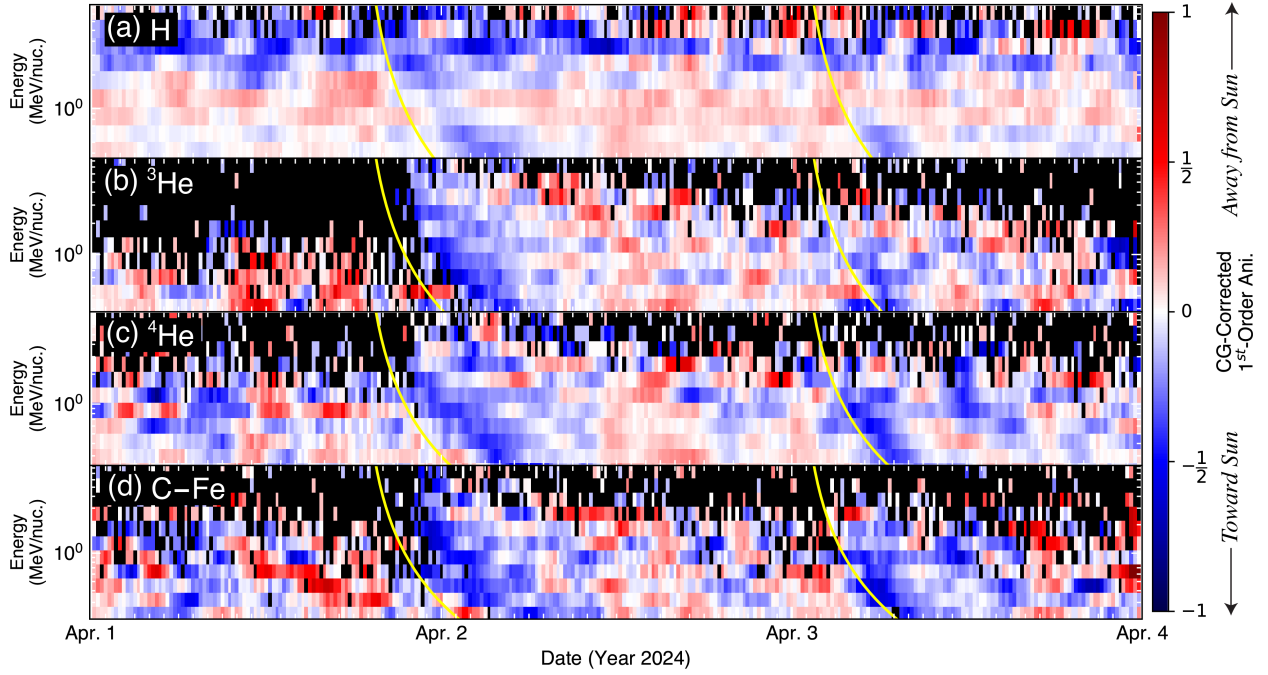

**Figure 5.** First-order anisotropy of (a) H, (b) $^3$He, (c) $^4$He, and (d) C – Fe during the two sunward streaming $^3$He-rich SEP events. Yellow curves indicate 1 AU dispersion lines. The persistent negative anisotropy in protons above 2 MeV/nuc is due to low proton detection efficiencies and should not be considered a permanent sunward streaming proton population. The 15-minute anisotropy measurements have been smoothed using a two-hour moving average.

We measure the first-order anisotropy of H, $^3$He, $^4$He, and C – Fe at each energy bin and at each 15-minute time interval, then we apply a two-hour moving average. Figure 5 presents the temporal profiles of the first-order anisotropy, with red and blue denoting positive and negative anisotropy, and white denoting isotropy. At the onset of both $^3$He-rich SEP events, we observe a strong negative anisotropy across the heavier ion species ($^3$He – Fe). Thus, during the early dispersive phase of these events, the bulk SEP flow is streaming sunward, consistent with Figure 4. The initial negative anisotropy follows the 1 AU velocity dispersion line. The SEPs then tend toward isotropic distributions on average, though there are brief periods with moderate positive and negative anisotropies. However, these periods are not consistent across species or energy and are likely statistical fluctuations. The negative anisotropy at event onset is not as apparent in H as in the heavier ions due to a persistent, elevated isotropic background from a previous gradual SEP event occurring days earlier. Again, the strong negative anisotropies observed in protons above 2 MeV/nuc are due to low detection efficiencies and should not be interpreted physically.

*3.3.2 Parker Solar Probe*

Using the same methods applied to SO/SIS (see §3.3.1), we derive the Compton-Getting corrected 2 – 8 MeV/nuc He energy spectra of event observed by PSP/LET. The results, shown in Figure 6, include spectra from all three LET telescopes. We find the energy spectra from PSP exhibits the same negative anisotropic feature seen by SO. During the dispersive phase, the LET-B and LET-C record consistently greater He intensities than LET-A, again indicating bulk SEP flow toward the Sun at event onset. Interestingly, the LET-B and LET-C intensities are mostly consistent with one another across the energy spectrum in the dispersive phase, suggesting the sunward streaming SEPs are not aligned with any particular telescope. Such an observation is consistent with SEPs streaming sunward along a near-radial IMF (see §3.2.2) with the LET-B and LET-C telescopes equally offset by 45º in the wake/ram direction, respectively. Finally, just as we observed at SO/SIS, all three PSP/LET telescopes observe the same He intensities during the decay phase indicative of an isotropic SEP distribution.

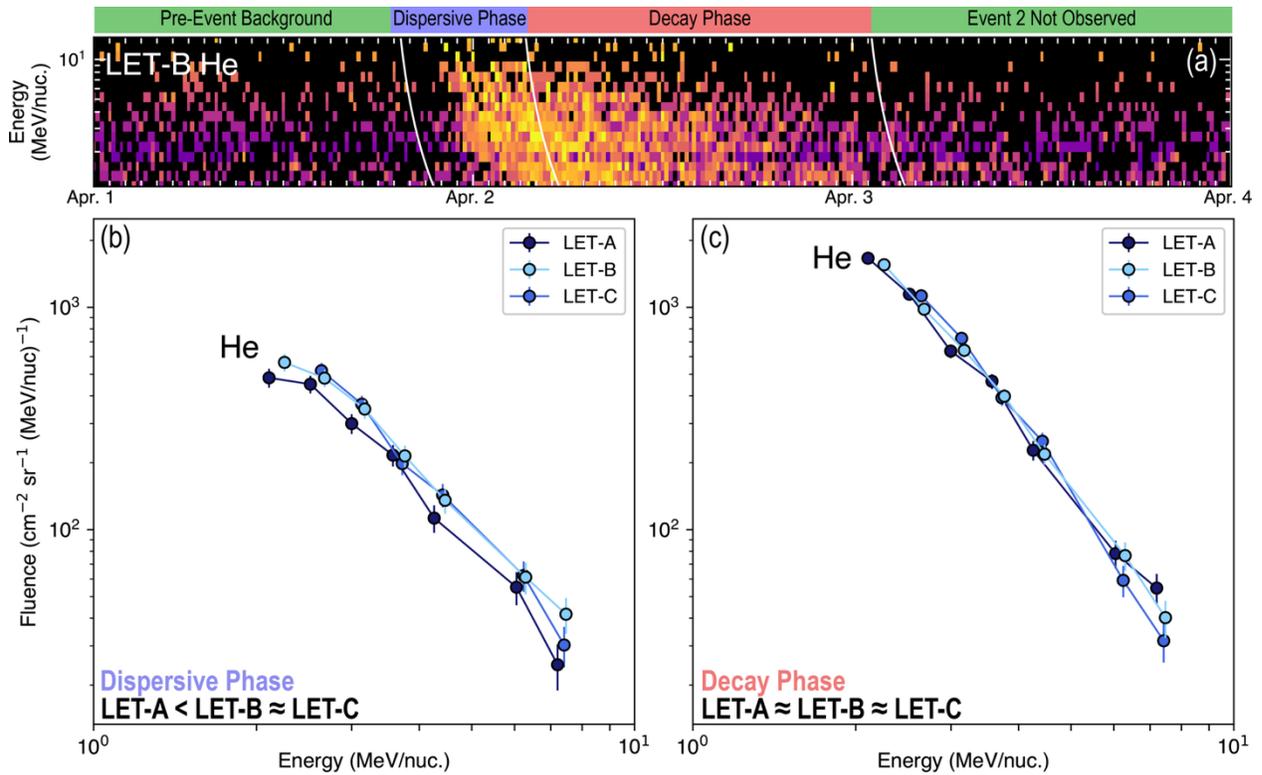

**Figure 6.** Same as Figure 4, but for He observations in the three PSP/LET telescopes. The LET-B (light blue) and LET-C (medium blue) telescopes observe greater intensities than LET-A (dark blue) during the dispersive phase of the April 1, 2024 $^3$He-rich SEP event, again indicating a bulk SEP sunward flow at event onset. During the decay phase, the intensities are equal at all telescopes, indicating an isotropic distribution.

To further quantify the SEP anisotropy, we calculate the first-order He anisotropy using equations 1 – 3 for each combination of telescope pairs (LET-A vs. B, A vs. C, B vs. C) separately, and we show the results of our analysis in Figure 7. Figures 7a show the first-order anisotropy of LET-A and LET-B. At the onset of the event, we observe the slight negative anisotropy (blue) seen in the energy spectra in Figure 6b. A similar feature can also be seen in the LET-A and LET-C anisotropy, shown in Figure 7b. Finally, Figure 7c shows the first-order anisotropy of LET-B and LET-C. Unlike with LET-A, the dispersive phase does not exhibit clear anisotropies indicating equal intensities observed by both antisunward facing telescopes. Compared with SO/SIS observations (see Figure 5), the SEP anisotropy at PSP is less pronounced. SEPs streaming into regions of greater IMF magnitude undergo adiabatic focusing, increasing their pitch angles closer to the Sun and potentially reducing the observed anisotropy. Alternatively, local increases in the turbulence near PSP can increase the efficacy of pitch angle scattering. As with SO/SIS, the brief periods of strong anisotropy during the decay phase are likely due to statistical fluctuations resulting from lower SEP intensities.

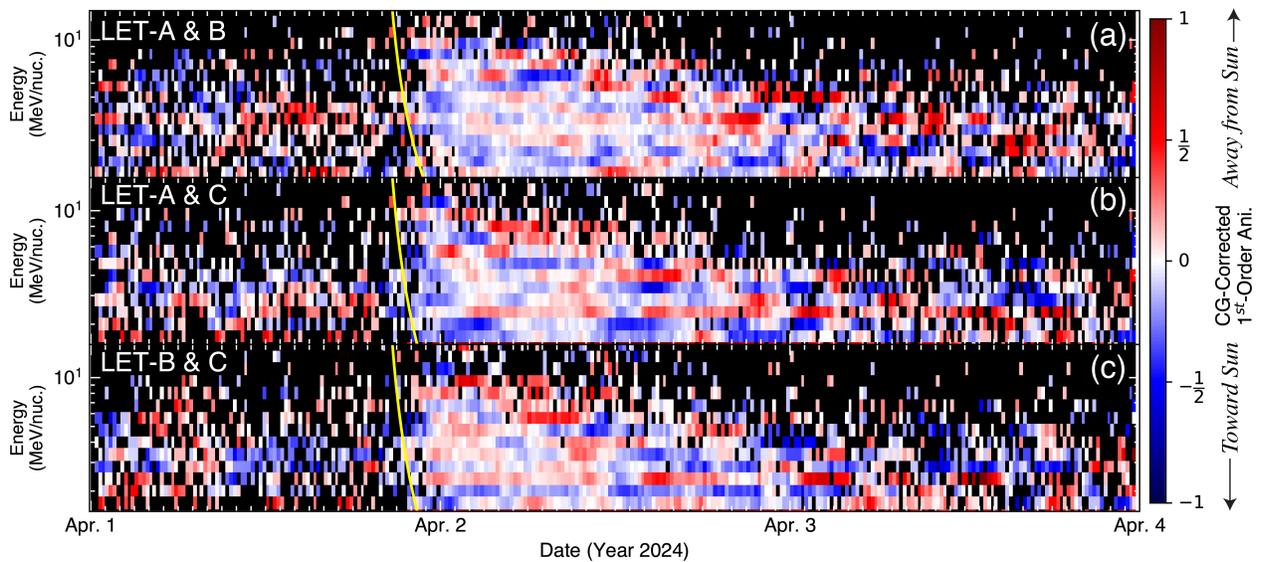

**Figure 7.** Same as Figure 5, but for PSP LET-A, LET-B, and LET-C He observations. The negative anisotropy is not as strong as in SO/SIS, possibly due to adiabatic focusing or increased pitch angle scattering as the $^3$He-rich SEPs propagate into increasingly stronger magnetic fields. LET-B and LET-C show nearly equal intensities during the dispersive phase.

*3.4 STEREO-A Observations*

STEREO-A is also nearly magnetically aligned with SO and PSP at this time, and it is equipped with the STEREO/LET sensor (Mewaldt et al. 2008) to measure He SEPs above 2 MeV/nuc. Figure 8 shows the hourly 2 – 10 MeV/nuc $^4$He intensity time profile observed by STEREO/LET along with the SO/SIS and PSP/LET intensities at 1.5 MeV/nuc at finer temporal resolution during event 1. The SEPs from event 1 are first observed by SO/SIS, then a short time later by PSP/LET. This particular order of arrival times is further evidence of a sunward streaming population, as SO is further from the Sun than PSP but is nevertheless the first to observe the $^3$He-rich SEP event. A short time later, a slight $^4$He enhancement (and $^3$He; M.E. Wiedenbeck, private communication) is observed by STEREO/LET. All three spacecraft exhibit similar SEP intensity decay profiles during the decay phase. The SEP enhancement observed by STEREO/LET at 1 AU is evidence to suggest some $^3$He-rich SEPs are not entirely redirected sunward but instead continue travelling outward to 1 AU. Thus, the particles appear to be widely distributed across and along the magnetic field lines from PSP (0.16 AU) to STEREO (1 AU), and the implications of the widespread distribution are discussed in §5.4.

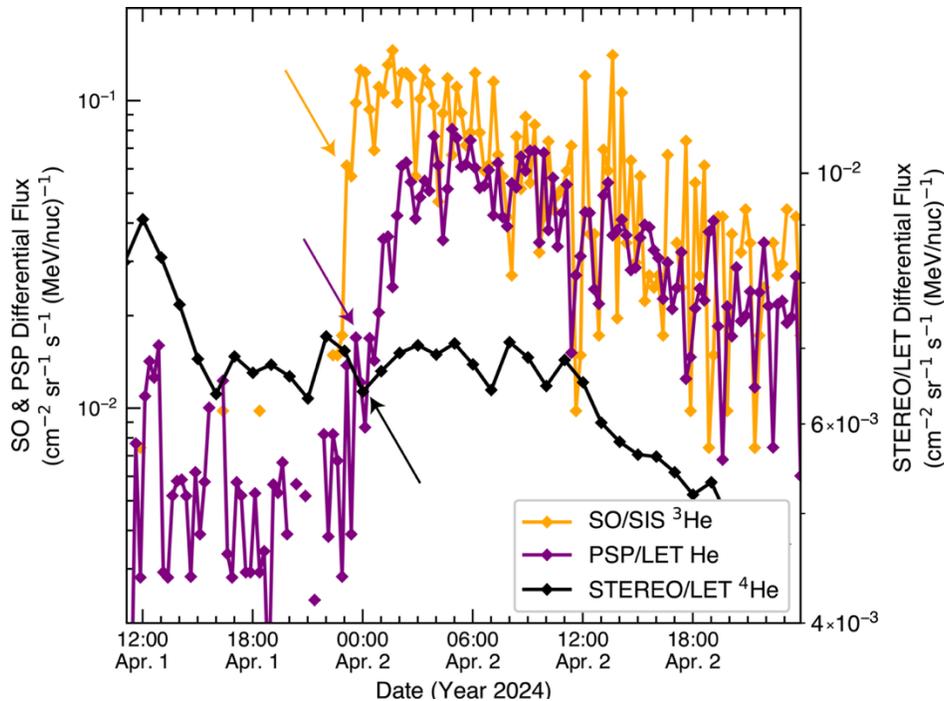

**Figure 8.** Intensity time profiles of SO/SIS $^3$He (orange) and PSP/LET He (purple) at 1.5 MeV/nuc as well as hourly STEREO/LET $^4$He at 2 – 10 MeV/nuc. Arrows indicate the time where the intensity enhacnements are first observed.

## 4. Remote Observations

Having established that the $^3$He-rich SEP events observed by Solar Orbiter and Parker Solar Probe exhibit sunward streaming anisotropies, we now identify their solar source regions using

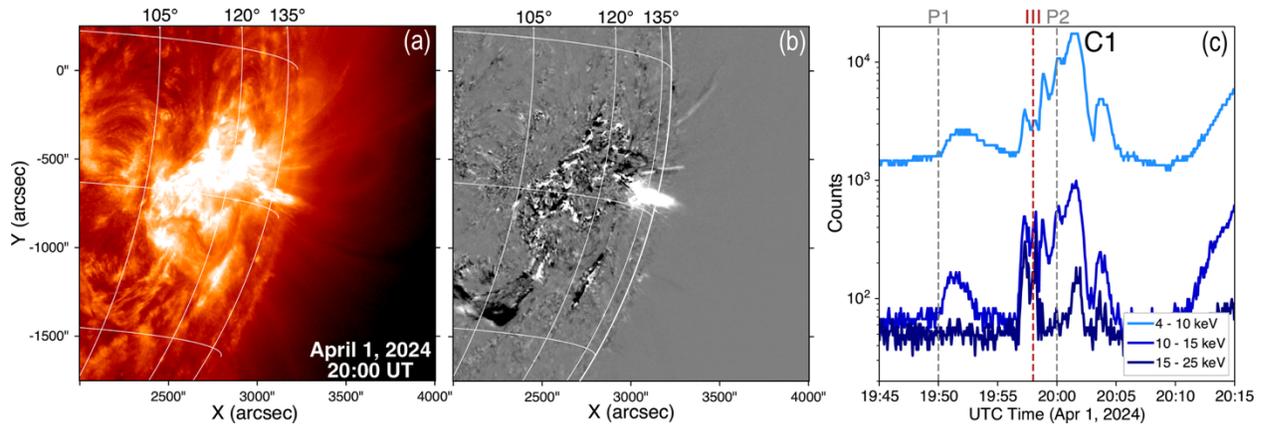

**Figure 9.** Solar source images of event 1 observed by both SO and PSP. (a) SO/EUI 304Å image following the type III burst onset. (b) 10-minute difference image. (c) SO/STIX X-ray counts vs. time. Vertical, dashed gray lines show the time at which the images in (b) were taken. Vertical, dashed red line indicates type III radio burst time. We include the estimated GOES flare class.

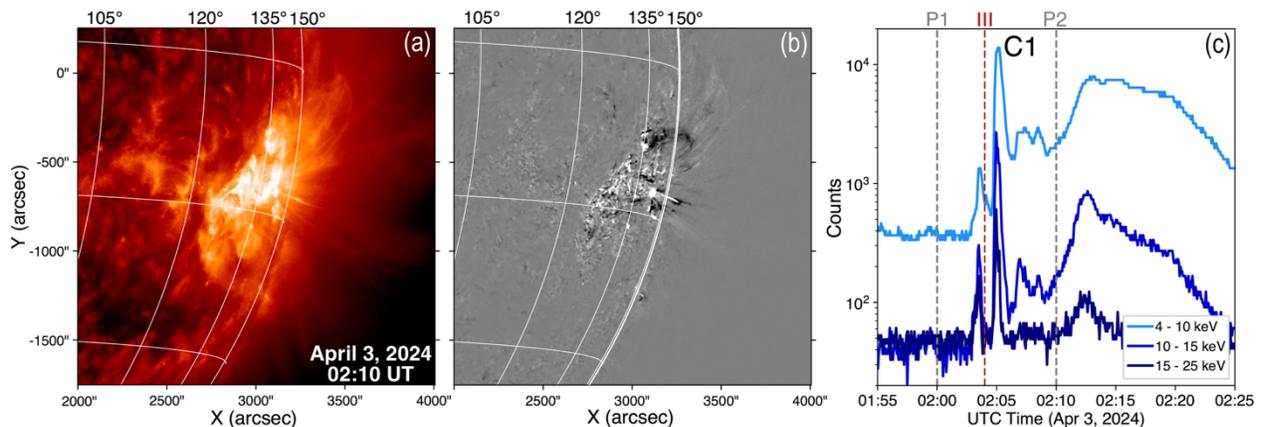

**Figure 10.** Same as Figure 9, but for event 2. The EUV jet is not as clear as event 1, likely because the EUV jet's peak brightness occurred at 02:05 UT, five minutes before (a) was taken.

SO/EUI and SO/STIX data near the onset times of the source type III radio bursts. Both events are associated with low C-class flares (estimated, see Xiao et al. 2023) originating from AR 13615. Figures 9 and 10 show SO/EUI and SO/STIX observations at the time of the first and second events, respectively. Panel (a) shows an SO/EUI 304Å image following the event 1's type III burst, and panel (b) shows the 10-minute difference image obtained by subtracting panel (a) from the image taken 10 minutes earlier, before the type III burst. The EUV jet associated with the first event is clearly captured, with the image taken within two minutes of the type III onset. The second event's EUV jet is not as clearly captured by EUV imaging, but the difference image in Figure 10b does show a brightening at the base of AR 13615 as well as a faint jet. The X-ray profile (Figure 10c) shows that the onset and decay of the second event occur between the two SO/EUI image timestamps, indicating only the tail end of the EUV jet was captured.

The solar source region, AR 13615, is located at 72°W and 76°W on the solar disk at the time of the events 1 and 2, respectively, from SO's perspective. While these values are typical for 1 AU observations (Reames 1999, Nitta et al. 2015, Bučík et al. 2016), they are unexpectedly further west for an observer at 0.3 AU. Using ballistic backmapping of the solar wind, we find AR 13615 is separated by ~50° from SO's source surface footpoint (see Figure 1). Unlike gradual SEP events, which are driven by widespread CME shocks, $^3$He-rich SEP events originate from a compact source in the lower solar corona and are spready typically 20 – 30° in heliospheric longitude (Ho et al. 2024). Given this context, a 50° separation between the SO footpoint and the solar source region indicates a major disturbance in the IMF from nominal Parker spiral conditions.

## 5. Results & Discussion

*5.1 CME Identification*

We explore the scenario outlined in Wimmer-Schweingruber et al. (2023), in which SEPs are injected onto the magnetic flux rope of a propagating ICME, causing elongated path lengths. Indeed, a well-placed ICME might explain the large measured path lengths, as broad ICMEs elongate magnetic field lines as they propagate through the heliosphere. However, as discussed in §3.2.1 and §3.2.3, *in situ* measurements from SO and PSP show no direct evidence of an ICME passage. Nevertheless, ICME redirection remains a plausible explanation. For the hypothesis to be viable, the ICME must meet the following criteria:

1) The ICME must propagate at sub-500 km/s speeds since no shock-accelerated SEP enhancements are observed.

2) The ICME must originate from or near AR 13615 to enable SEP injection along the ICME front.

3) The estimated path along the ICME at SEP arrival time must be ~1 AU to match the observed dispersion profiles at both SO and PSP.

Using the SOHO LASCO CME Catalog (Gopalswamy et al. 2024), we searched for slow CMEs originating near AR 13615 within three days prior the $^3$He-rich SEP events. We identified a potential candidate - a CME associated with a GOES M9-class X-ray flare originating on March 30, 2024 at 21:04 UT, ~48 hours before the first arriving $^3$He-rich SEPs. Despite the large class X-ray flare, the associated CME is weak, with near-Sun speeds well below 500 km/s and an plane-of-sky angular width of $\theta_{CME} = 71°$.

*5.2 CME Properties*

To determine the speed of the ICME, we track the height of its bright front in six successive LASCO C2 12-minute difference images (see Figure 11). At each 3° angle bin, the CME front is found by locating the height of peak brightness and then stepping outward to where the brightness falls to half that value (the red crosses in Figure 11). To obtain the total CME height, we perform least squares regression on the CME front using an ellipse function of the polar form:

$$r(\theta) = \frac{A(1-e^2)}{1-e\cos(\theta)}, 0 \leq \theta \leq 2\pi \tag{4}$$

where $r(\theta)$ is distance at each angle, $A$ is the semi-major axis, and $e$ is eccentricity. We then fit both first-order and second-order polynomials to the CME height (distance from solar surface to CME nose) as a function of time to obtain CME speeds. From the results of the first-order polynomial fit, we obtain a linear CME speed of 281 ± 6 km/s. However, the residuals of the first-order fit (top panel of Figure 12) suggest the presence of residual acceleration (Gopalswamy et al. 2000, Bein et al. 2011). Fitting the profile to a second-order fit yields a residual acceleration of 32 ± 12 m/s$^2$, within typical values (Zhang et al. 2006). Ultimately, we choose the second-order speed to be the speed at 3.5 solar radii, since above this height the fitting becomes difficult due to poor data. At 3.5 solar radii, we obtain a second-order best-fit speed of 368 ± 26 km/s. The best-fit CME front eccentricity remains roughly constant, centered at *e* = 0.62 during this period.

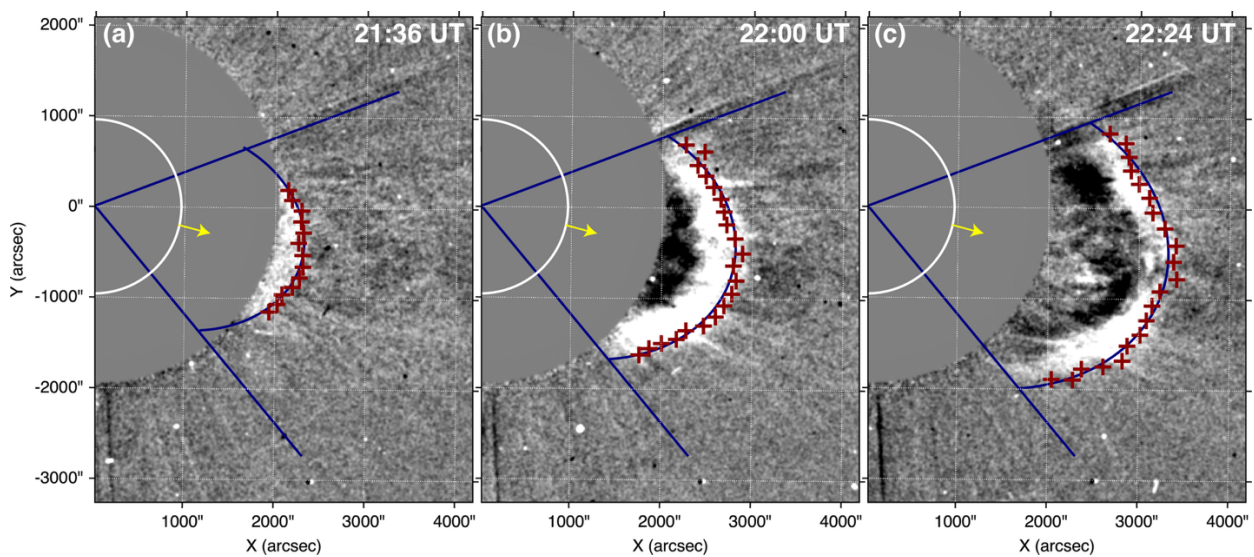

**Figure 11.** (a) LASCO C2 12-minute difference of the candidate CME erupting from AR 13615 following an M9 solar flare on March 30, 2024 21:04 UT. The CME height is tracked by fitting an ellipse (blue arc) to the CME fronts at each 3° angle bin (red crosses). (b – c) same as (a) but at later times.

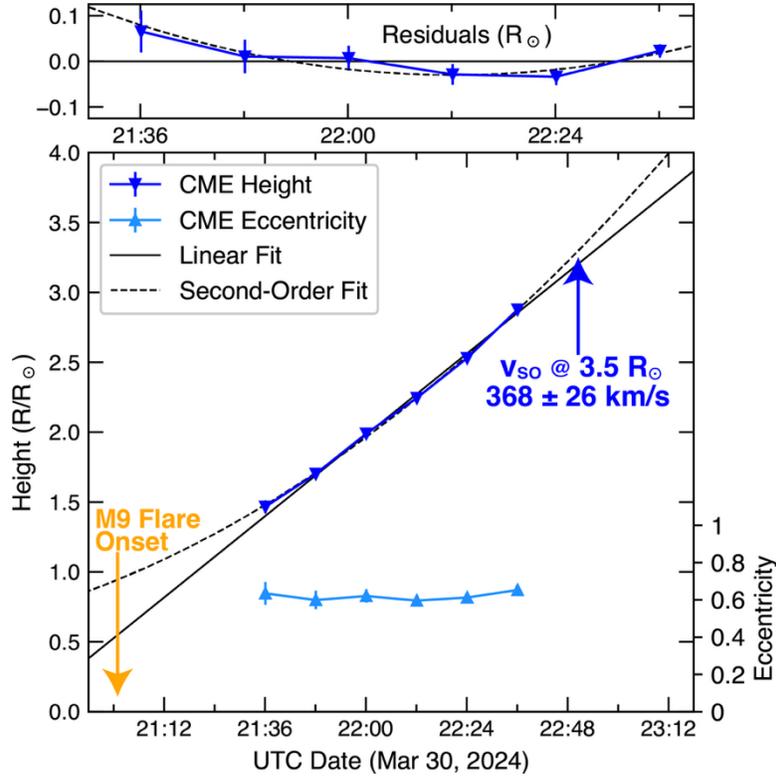

**Figure 12.** Results of the elliptical fits to the CME front. The free parameters are semi-major axis, which we have converted to a height above the photosphere (dark blue triangles), as well as an eccentricity (light blue triangles). We then fit a first-order (solid black line) and second-order (dashed black line) polynomial the height versus time profile. Residuals to the fits are shown at the top of the figure.

*5.3 Modelled SEP Path Lengths*

To estimate the SEP path lengths, we assume the near-sun CME maintains a constant speed of 368 ± 26 km/s, eccentricity of *e* = 0.62, and we assume the measured plane-of-sky width ($\theta_{CME} = 71°$) is also the ecliptic width. Using the time delay between the March 30, 2024 M9-class flare and first arriving $^3$He-rich SEPs, we propagate this ICME through the heliosphere to model the SEP path from the Sun as the particles travel along the draped magnetic field lines. An illustration of the scenario is shown in Figure 13, with the SEP travel path shown in blue. The SEPs are injected onto a nominal Parker spiral field, but the field drapes around the propagating ICME, causing the SEPs to be redirected to the ICME limb. The SEPs then travel around the ICME before arriving at SO and later PSP. Each blue arrow in Figure 13 represents a 0.1 AU step along the travel path. Using our rough model of the SEP travel path, we find a total path length ranging between 0.76 – 0.95 AU at SO and 0.94 – 1.1 AU at PSP for the first $^3$He-rich SEP event.

Our modeled path lengths are approximately 15% (23%) less than the observed SO (PSP) path length of 0.99 AU (1.33 AU) derived from the 1.5 MeV/nuc ion arrival times (see §3.2.3). The reduced path lengths from our model may arise from assumptions such as constant CME speed

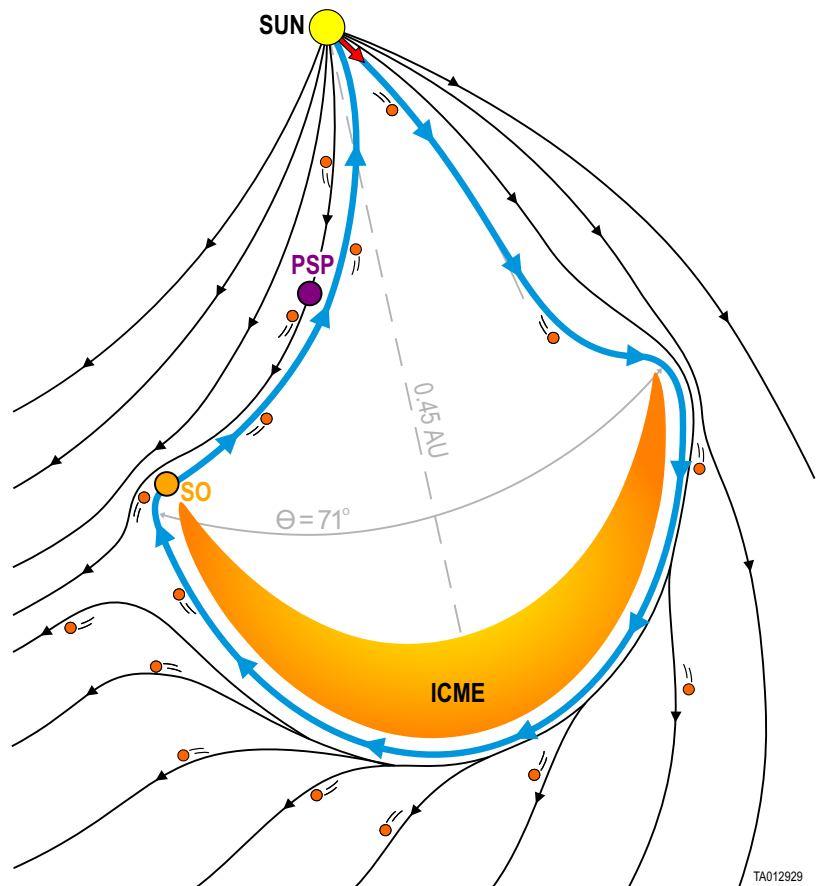

**Figure 13.** Illustration of the ICME at the time of the first $^3$He-rich SEP event. The SEPs (small red circles) follow the thick blue curve around the ICME, and each blue arrow represents an additional 0.1 AU travelled. The majority of SEPs return to the Sun to be a pre-accelerated seed population for subsequent SEP events, but some continue travelling outward, causing a widespread distribution of $^3$He-rich SEPs.

Sunward Streaming $^3$He-rich SEPs 24

and CME symmetry as well as our assumption that the SEPs travel along the same plane, whereas Wimmer-Schweingruber et al. (2023) show out-of-ecliptic redirection is also possible. Furthermore, we do not account for small-scale meandering effects that tend to elongate path lengths (Chhiber et al. 2021). While the underestimate suggests our model is overly simplistic, the results still show that the observed sunward streaming $^3$He-rich SEPs can be plausibly explained by an SEP injection onto elongated magnetic field lines draped around an inner heliospheric ICME.

*5.4 Implications of Results*

The results of this case study have implications on three heliospheric phenomena: the cause of unusually long SEP path lengths, the role of $^3$He-rich SEPs in supplying seed particle populations for gradual SEP events, and the nature of broadly distributed $^3$He-rich SEP events.

*Unusually Long Path Lengths:* In this work, we examine two $^3$He-rich SEP events with exceptionally long path lengths. As in Wimmer-Schweingruber et al. (2023), our analysis points toward the redirection of SEPs around an inner heliospheric ICME. During solar max, nearly 50% of solar wind flow at 1 AU is of ICME origin (Kilpua et al. 2017), suggesting ICME redirection of SEPs in the inner heliosphere may be a common occurrence. Continued SEP observations within about 0.4 AU will address the question of frequency.

*$^3$He-rich Seed Population:* We provide the first multi-point observation of sunward streaming $^3$He-rich SEP events. Our results indicate these SEPs are released from the Sun, travel outward for several tenths of an AU, and ultimately return to the solar corona, where they presumably remain for some time due to adiabatic focusing and magnetic mirroring. In this scenario, an entire $^3$He-rich SEP event's worth of seed material can populate the coronal environment above an AR primed for reacceleration. Subsequent gradual SEP events may then be

heavily enriched with $^3$He-rich material, but there will be no remote signatures (i.e., EUV jets) to indicate a simultaneous injection of $^3$He-rich SEPs (see Bučík et al. 2025a). However, it is still not clear how frequently this phenomenon occurs.

*Widespread $^3$He-rich SEP events:* Observations of widespread $^3$He-rich SEP events have been recorded, with some events spanning nearly 180° in heliospheric longitude, and it remains unclear what mechanism(s) enable the SEPs to achieve such broad distributions. Our observations provide a potential explanation: if $^3$He-rich SEPs diffuse onto many different field lines as they propagate along an ICME front, then they will be distributed over the full width of the ICME. In this case, the broad extent of the ICME acts as a dispersive mechanism, allowing $^3$He-rich SEPs to be distributed over a large range of heliospheric longitudes. Importantly, the ICME in our study existed for two days prior to the $^3$He-rich SEP events. Thus, the $^3$He-rich SEP events themselves need not be associated with CMEs for them to exhibit widespread characteristics. The small $^3$He and $^4$He enhancements observed by STEREO at 1 AU are evidence that not all SEPs were redirected toward the Sun. Still, other proposed explanations for widespread $^3$He-rich SEP events, such as large-scale coronal waves (Bučík et al. 2016) and complex configurations of the separatrix-web (S-Web; Scott et al. 2018), have not been explored in this work. Future investigations of widespread $^3$He-rich SEP events should survey for slow, inner heliospheric ICMEs as a potential dispersive mechanism for SEPs.

However, several open questions remain, particularly regarding SEP transport processes along pre-existing ICMEs. While the ICME hypothesis explains the extended SEP path lengths and sunward streaming SEPs observed during the events, neither SO nor PSP detects definitive *in situ* ICME signatures. The candidate ICME's angular width ($\theta_{CME} = 71°$) places both SO and PSP just at the edge of the ICME limb (see Figure 13), so it may be the case that the two spacecraft

"just missed" the ICME. On what field line(s) were the $^3$He-rich SEPs injected? Were they propagating along closed field lines and diffused onto open field lines? Furthermore, the exact path the SEPs travelled is unknown. The second $^3$He-rich SEP event's path length is 16% shorter than the first event's path length, but the ICME has continued to propagate outward, suggesting the path length ought to be much longer. Did the SEPs from the second event follow a different path than the first event? Had the global magnetic configuration in the inner heliosphere mostly reverted to nominal conditions by the onset of the second event?

## 6. Conclusion

In this work, we investigate two $^3$He-rich SEP events originating from AR 13615 between April 1 – 4, 2024 that were observed by SO/SIS at a radial distance of 0.3 AU. At SO/SIS, the two SEP events exhibit: (1) negative anisotropies, meaning the bulk SEP flow is streaming toward the Sun rather than away from it, and (2) anomalously large path lengths of 0.99 AU and 0.83 AU, exceeding 2 – 3 times the expected path length along the nominal Parker spiral field. Combining these two observations, we suggest the $^3$He-rich SEPs were injected into the heliosphere, but were redirected back toward the Sun by an inner heliospheric ICME. PSP, located at 0.16 AU and nearly magnetically aligned with SO, also observes the first of the two events. PSP/LET analysis confirms the SO/SIS observations, as the first arriving $^3$He-rich SEPs observed by PSP also exhibit negative anisotropies, and the calculated path length based on ion arrival times at PSP is 1.33 AU, over 8 times greater than the nominal Parker spiral path length. When comparing the SEP arrival times at the two inner heliospheric spacecraft, we find that the SEPs arrive first at SO at 0.3 AU, then later at PSP at 0.16 AU, again indicative of a sunward flow of SEPs over at least one-tenth of an AU.

We consider the scenario described in Wimmer-Schweingruber et al. (2023), in which a $^3$He-rich SEP event observed by SO exhibits an unusually long path length during the passage of a clear ICME flux rope. Because there are no clear indicators of an ICME in the *in situ* plasma parameters at SO and PSP, we infer candidate CME characteristics from remote sensing observations. Using SOHO/LASCO, we identify a slow ICME originating from AR 13615 on March 30, 2024 at 21:04 UT, ~48 hours prior to the first arriving $^3$He-rich SEPs. The CME exhibited a near-Sun speed of 368 ± 26 km/s and an angular width of 71°. Based on these parameters, we propagate the ICME through the heliosphere to obtain an estimate of its size and location at the time of the first $^3$He-rich SEP event, and we provide a rough model of the SEP path around the ICME front. The calculated path lengths of 0.76 – 0.95 AU at SO and 0.94 – 1.1 AU at PSP are shorter than, but in reasonable agreement with, the measured values of 0.99 ± 0.05 AU and 1.33 ± 0.05 AU at SO and PSP, respectively.

This work provides results for the first multi-spacecraft observation of sunward streaming $^3$He-rich SEP events. Though it remains an open question how often this phenomenon occurs, it provides a mechanism for localized regions in the solar corona to be heavily enriched with $^3$He-rich seed material, primed for reacceleration by subsequent gradual SEP events. Furthermore, SEP diffusion along the ICME front may explain why some $^3$He-rich SEP events are observed to be broadly distributed across heliospheric longitude. Future multi-spacecraft observations of this phenomenon will help refine the interpretations presented here. We are hopeful for new observations as inner heliospheric observatories continue to measure the SEP environment near the Sun.

**7. Acknowledgements**

The work performed in this manuscript is done primarily using the support of NASA Contract 80GSFC25CA035. S. T. Hart and R. Bučík also thank M. E. Wiedenbeck for his useful discussions under NASA grant 80NSSC21K1316. M. A. Dayeh also acknowledges partial support from NASA award 80NSSC25K7687. The Suprathermal Ion Spectrograph (SIS) is a European facility instrument funded by ESA under contract number SOL.ASTR.CON.00004. We thank ESA and NASA for their support of the Solar Orbiter and other missions whose data were used in this letter. Solar Orbiter post-launch work at JHU/APL is supported by NASA contract NNN06AA01C.

**Appendix A. Coronal Hole Observation**

We describe coronal hole boundaries (CHs) as a common source region for $^3$He-rich SEP events, particularly when they are neighboring solar active regions (ARs). Because the $^3$He-rich SEP events analyzed in this work occur near the western limb with respect to SO and behind the limb with respect to ACE and STEREO-A, the CH neighboring AR 13615 (see Figure 1) cannot be easily seen in the solar source images at the time of the two $^3$He-rich SEP events. The existence of the CH is not pertinent to the analysis, so we provide an image of it in Figure A1. Figure A1 shows an image taken by the Solar Dynamics Observatory Atmospheric Imaging Assembly (SDO/AIA; Pesnell et al. 2012, Lemen et al. 2012) at 211Å. CHs are characterized as patches on the solar disk that are exceptionally dim at 193Å and 211Å because they are cooler and less dense, which result in less Fe ionization to high ionization states whose emissions overlap with 193Å and 211Å (Del Zanna et al. 2015). CHs are also characterized by their open magnetic field lines required for $^3$He-rich SEPs to escape into interplanetary space.

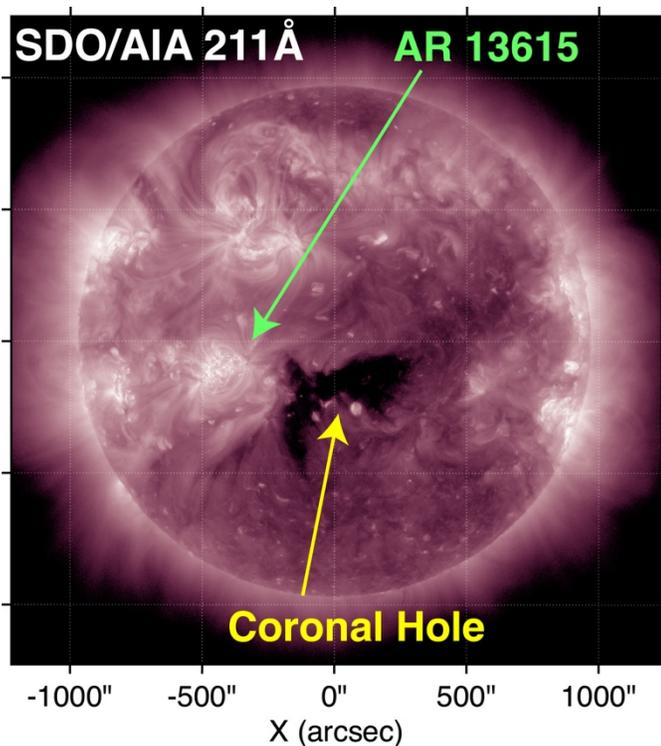

**Figure A1.** SDO/AIA 211Å image on March 25, 2024 00:00 UT. AR 13615 sources the $^3$He-rich SEP events and is neighboring a low-latitude coronal hole. The $^3$He-rich SEP events originate at the leading edge of AR 13615, strongly indicating reconnection along the AR – CH boundary.

# 8. References

Acton, C. H., 1996, PSS, 44, 65
Allen, R. C., Ho, G. C., Mason, G. M., et al. 2025, ApJ, 982, 160
Bale, S. D., Goetz, K., Harvey, P. R., et al. 2016, SSRv, 204, 49
Bein, B. M., Berkebile-Stoiser, S., Veronig, A. M., et al. 2011, ApJ, 738, 191
Brueckner, G. E., Howard, R. A., Koomen, M. J., et al. 1995, SoPh, 162, 357
Bučík, R., Innes, D. E., Mason, G. M., Wiedenbeck, M. E. 2016, ApJ, 833, 63
Bučík, R., Wiedenbeck, M. E., Mason, G. M., et al. 2018, ApJL, 869, L21
Bučík, R. 2020, SSRv, 216, 24
Bučík, R., Mason, G. M., Mulay, S., et al. 2025a, ApJ, 981, 178
Bučík, R., Hart, S. T., Dayeh, M. A., et al. 2025b, ApJ, 994, 252
Carcaboso, F., Gómez-Herrero, R., Espinosa Lara, F., et al. 2020, A&A, 635, A79
Chhiber, R., Matthaues, W. H., Cohen, C. M. S., et al. 2021, A&A, 650, A26
Compton, A. H., Getting, I. A. 1935, Physical Review, 47, 817
Dayeh, M. A, McComas, D. J., Allegrini, F., et al. 2012, ApJ, 749, 50
Domingo, V., Fleck, B., & Poland, A. I. 1995, SoPh, 162, 1
Dresing, N., Gómez-Herrero, R., Heber, B., et al. 2014, A&A, 567, A27
Fitzmaurice, A., Drake, J. F., Swisdak, M. 2024, ApJ, 964, 97
Fox, N. J., Velli, M. C., Bale, S. D., et al. 2016, SSRv, 204, 7
Gopalswamy, N., Lara, A., Lepping, R. P., et al. 2000, GRL, 27, 145
Gopalswamy, N., Michalek, G., Yashiro, S., et al. 2024, arXiv:2407.04165
Hart, S. T., Dayeh, M. A., Bučík, R., et al. 2022, ApJS, 263, 22
Hart, S. T., Dayeh, M. A., Bučík, R., et al. 2024, ApJ, 974, 220
Henderson, S. A., Filwett, R. J., Lee, C. O., et al. 2025, ApJ, 992, 34
Ho, G. C., Mason, G. M., Allen, R. C., et al. 2022, FRASS, 9, 939799
Ho, G. C., Mason, G. M., Allen, R. C., et al. 2024, ApJ, 974, 68
Horbury, T. S., O'Brien, H., Carrasco Blazquez, I., et al. 2020, A&A, 642, A9
Ipavich, F. M. 1974, GRL, 1, 149
Kahler, S., Reames, D. V., Sheeley, N. R., Jr. et al. 1985, ApJ, 290, 742
Kasper, J. C., Abiad, R., Austin, G., et al. 2016, SSRv, 204, 131
Kilpua, E., Koskinen, H. E. J., Pulkkinen, T. I., 2017, LRSP, 14, 5
Kouloumvakos, A., Mason, G. M., Ho, G. C., et al. 2025, ApJ, 983, 67
Laitinen, T., & Dalla, S. 2019, ApJ, 889, 222
Laming, J. M., & Kuroda, N. 2023, ApJ, 951, 86
Mason, G. M., Wiedenbeck, M. E., Miller, J. A., et al. 2002, ApJ, 574, 1039
Mason, G. M., Mazur, J. E., Dwyer, J. R., et al. 2004, ApJ, 606, 555
Mason, G. M. 2007, SSRv, 130, 231
Mason, G. M. 2021, Ho, G. C., Allen, R. C. et al. 2021, A&A, 656, L1
Mason, G. M., Roth, I., Nitta, N. V., et al. 2023a, ApJ, 957, 112
Mason, G. M., Nitta, N. V., Bučík, R., et al. 2023b, A&A, 669, L16
McComas, D. J., Alexander, N., Angold, N., et al. 2016, SSRv, 204, 187
Mewaldt, R., Cohen, C. M. S., Cook, W., et al. 2008, SSRv, 136, 285
Müller, D., Cyr, O. C. S., Zouganelis, I., Gilbert, H. E., & Marsden, R. 2020, A&A, 642, A1
Nitta, N. V., Mason, G. M., Wang, L., Cohen, C. M. S., & Wiedenbeck, M. E. 2015, ApJ, 806, 235
Owen, C. J., Bruno, R., Livi, S., et al. 2020, A&A, 642, A16
Reames, D. V., & Stone, R.G. 1986, ApJ, 308, 902

Reames, D. V. 1999, SSRv, 90, 413
Reames, D. V. 2021, FRASS, 8, 164
Rochus, P., Auchère, F., Berghmans, D., et al. 2020, A&A, 642, A8
Rodríguez-Pacheco, J., Wimmer-Schweingruber, R. F., Mason, G. M., et al. 2020, A&A, 642, A7
Scott, R. B., Pontin, D. I., Yeates, A. R., Wyper, P. F., Higginson, A. K. 2018, ApJ, 869, 60
Shimojo, M., & Shibata, K. 2000, ApJ, 542, 1100
Shmies, A. A., Dayeh, M. A., Bučík, R. et al. 2026, in prep
Starkey, M. J., Dayeh, M. A., Desai, M. I., et al. 2024, ApJ, 962, 160
Wang, Y.-M., Pick, M., & Mason, G. M. 2006, ApJ, 639, 495
Wei, W., Lee, C. O., Dresing, N., et al. 2024, ApJL, 973, L52
Wiedenbeck, M. E., Mason, G. M., Cohen, C. M. S., et al. 2013, ApJ, 762, 54
Wimmer-Schweingruber, R. F., Janitzek, N. P., Pacheco, D., et al. 2021, A&A, 656, A22
Wimmer-Schweingruber, R. F., Berger, L., Kollhoff, A., et al. 2023, A&A, 678, A98
Xiao, H., Maloney, S., Krucker, S., et al. 2023, A&A, 673, A142
Xu, Z. G., Cohen, C. M. S., Leske, R. A., et al. 2024, ApJ, 976, L3
Zirnstein, E. J., Dayeh, M. A., Heerikhuisen, J., McComas, D. J., & Swaczyna, P. 2021, ApJS, 252, 26
Zhang, J., & Dere, K. P. 2006, ApJ, 649, 1100Sunward Streaming $^3$He-rich SEPs    32